\documentclass[aps,prl,reprint]{revtex4-1}
\usepackage{graphicx}
\usepackage{amssymb}
\usepackage{amsmath}
\usepackage{bm}

\begin{document}

\title{Realization of a cross-linked chiral ladder with neutral fermions in an optical lattice by orbital-momentum coupling}

\author{Jin Hyoun Kang}
\author{Jeong Ho Han}
\author{Y. Shin}\email{yishin@snu.ac.kr}
\affiliation{Department of Physics and Astronomy, and Institute of Applied Physics, Seoul National University, Seoul 08826, Korea
\\Center for Correlated Electron Systems, Institute for Basic Science, Seoul 08826, Korea}

\begin{abstract}
We report the experimental realization of a cross-linked chiral ladder with ultracold fermionic atoms in an optical lattice. In the ladder, the legs are formed by the orbital states of the optical lattice and the complex inter-leg links are generated by the orbital-changing Raman transitions that are driven by a moving lattice potential superimposed onto the optical lattice. The effective magnetic flux per ladder plaquette is tuned by the spatial periodicity of the moving lattice, and the chiral currents are observed from the asymmetric momentum distributions of the orbitals. The effect of the complex cross links is demonstrated in quench dynamics by measuring the momentum dependence of the inter-orbital coupling strength. We discuss the topological phase transition of the chiral ladder system for the variations of the complex cross links.

\end{abstract}

\maketitle

Topological states of matter represent one of the frontiers of modern condensed matter physics~\cite{Wen90,Hasan10}. Featuring tunable artificial gauge fields and spin-orbit coupling~\cite{Goldman14}, ultracold atoms in optical lattices provide a versatile platform to realize topological states and study their phase transitions in a clean and well-controlled manner~\cite{Jaksch05,Bloch08}. The Hofstadter-Harper model, which is the paradigmatic example of a topological Chern insulator, was realized in two-dimensional (2D) optical lattices using laser-assisted tunneling effects~\cite{Aidelsburger13,Miyake13}; the Haldane model was demonstrated in a hexagonal optical lattice by activating complex next-nearest-neighbor (NNN) hopping with lattice shaking~\cite{Jotzu14}. The high tunability of the experimental parameters may enable one to explore a broad range of topological states even beyond the conventional Altland-Zirnbauer classification~\cite{Altland97,Song18}. 

In recent optical-lattice experiments, an interesting framework was introduced to realize chiral ladder systems with one-dimensional (1D) optical lattices, where the internal atomic degrees of freedom, such as hyperfine spin, are taken as a finite synthetic dimension and an artificial gauge field is engineered by laser-induced couplings between the internal states~\cite{Boada12,Celi14}. The synthetic ladder systems are highlighted by their edges, which are intrinsically sharp and can be detected by internal-state-selective imaging, thus allowing direct observation of the chiral edge currents in the systems~\cite{Stuhl15,Mancini15,Livi16}. Ladders with complex hopping amplitudes have been discussed as a minimal model for 1D topological matter~\cite{Creutz99} and also as a quasi-1D version of the Hofstadter problem for studying the edge-mode states in 2D topological insulators~\cite{Hugel14}. In particular, it is well recognized that NNN hopping, i.e., diagonal cross links between the legs, are responsible for the emergence of topologically non-trivial phases in the ladder system subject to a magnetic field~\cite{Hugel14,Junemann17,Sun17}. Therefore, extensive control of the complex inter-leg links is highly desirable in synthetic ladder experiments.

In this paper, we report the experimental demonstration of a synthetic ladder scheme using the orbital degree of freedom of the optical lattice system. The legs of the ladder are formed by the orbital states, and the inter-leg hoppings are generated by orbital-changing two-photon Raman transitions that are resonantly driven by a moving lattice potential.  The complex hoppling amplitude is spatially modulated, giving rise to an effective magnetic flux $\Phi$ per ladder plaquette, which we demonstrate by observing the corresponding chiral currents of the orbital states. The key feature of our orbital-based ladder system is that the cross inter-leg links are significantly strong due to the favorable condition for the spatial overlap of the orbital wave functions. The complex cross link effect is manifested in the momentum dependence of the inter-orbital coupling strength, which we directly demonstrate via momentum-resolving analysis of the quench dynamics of the ladder system. Finally, we discuss the topological phase transition, which can occur in the system by further controlling the cross link, possibly via tailoring the orbital wave functions. Our results present a new perspective for studies of topological phases with optical lattice systems using the orbital degree of freedom~\cite{Li13,Zhang14,Khamehchi15}.

\begin{figure}[t]
	\includegraphics[width=8.4cm]{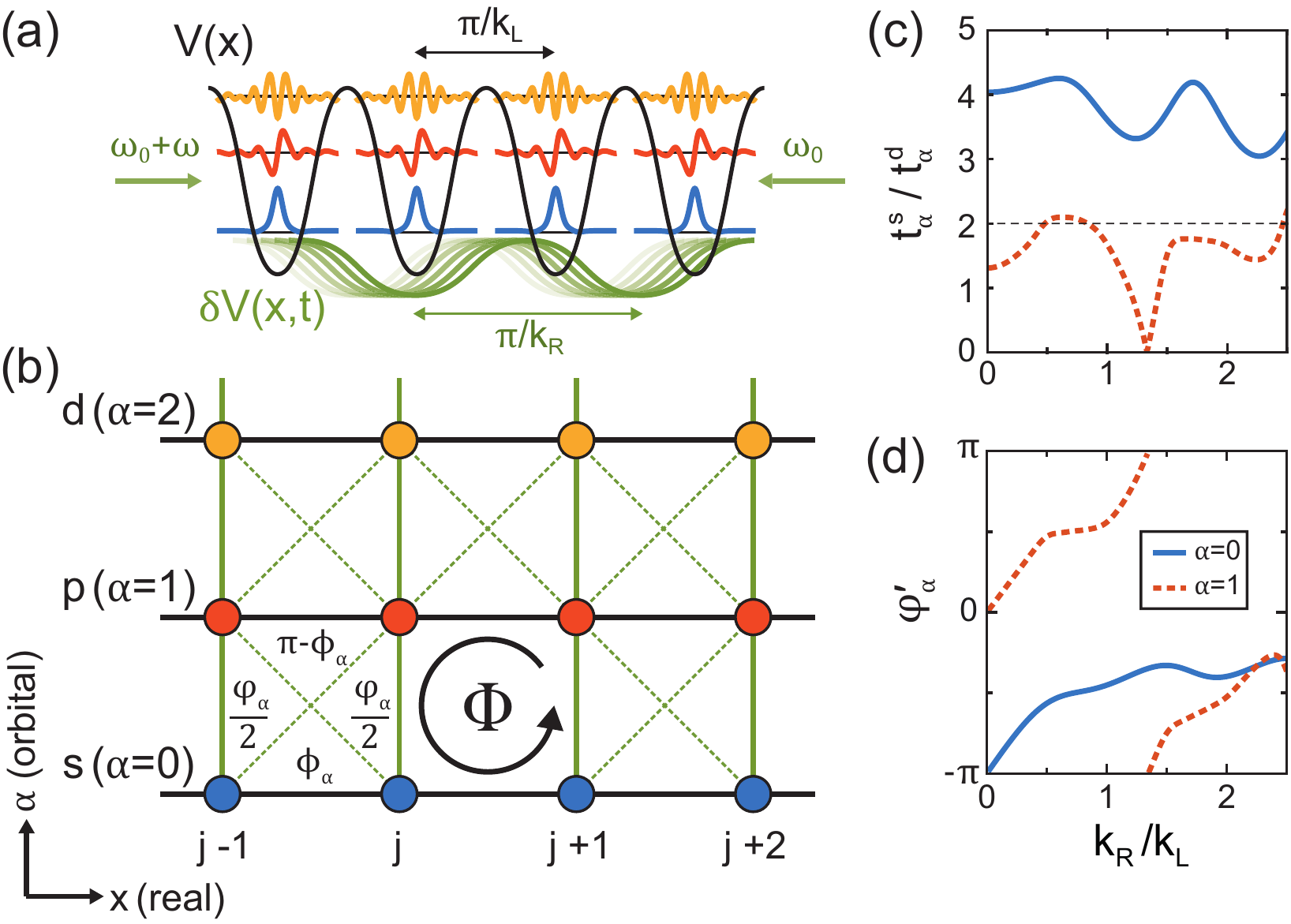}
	\caption{Orbital-momentum coupling in an optical lattice. (a) Schematic of the experimental setup. Atoms are in a stationary 1D optical lattice potential $V(x)$, with lattice constant $\pi/k_L$, and driven by a moving lattice potential $\delta V(x,t)$, with lattice constant $\pi/k_R$. The moving lattice induces two-photon Raman transitions between the orbital states of the stationary lattice. (b) Ladder description of the system. The orbitals constitute the ladder legs along the real $x$ dimension and the Raman coupling provides the inter-leg links. The complex coupling amplitude is spatially modulated, resulting in an effective magnetic flux per plaquette, $\Phi = 2\pi(k_{R}/k_{L})+\pi$. The sub-plaquette flux distribution is assigned in accordance with the cross links indicated by the diagonal dashed green lines. (c) Amplitude ratio for the direct to diagonal hopping, $t^{s}_\alpha/t^d_\alpha$ and (d) their phase difference $\varphi_{\alpha}'$ as functions of $k_R/k_L$.} 
\end{figure}

We consider a system of non-interacting spinless fermions in an 1D lattice potential $V(x)=V_L \cos^2 (k_L x)$, where the fermions are perturbed by a moving lattice potential, $\delta V(x,t)=V_R \cos^2 (k_R x -\frac{\omega t}{2})$ [Fig.~1(a)].  The Hilbert space of the system is spanned by the Wannier states $\{|j,\alpha\rangle\}$ of the stationary lattice potential, where $j$  and $\alpha$ are the indices for the lattice site and orbital, respectively. Regarding the orbital degree of freedom as a virtual dimension orthogonal to the real lattice dimension, the system can be viewed as a synthetic 2D lattice system that has a sharp edge formed by the $s$ orbital, as depicted in Fig.~1(b). When the modulation frequency $\omega$ of the moving lattice potential is close to a band gap, orbital-changing Raman transitions are resonantly driven, realizing hoppings for the synthetic dimension. 

In a multi-band tight-binding description~\cite{Dutta15}, using the rotating wave approximation, the system's Hamiltonian is given by~\cite{SI}
\begin{align}
H=&\sum_{j,\alpha} \bigg[(\epsilon_{\alpha}-\alpha \hbar \omega) c_{j,\alpha}^{\dagger}c_{j,\alpha} +\big( (-1)^{\alpha+1}t_{\alpha}^r c_{j,\alpha}^{\dagger}c_{j+1,\alpha} + \textrm{h.c.}\big) \nonumber \\
&+\frac{1}{2}\big( t_{\alpha}^s e^{-i \varphi j} c_{j,\alpha}^{\dagger}c_{j,\alpha+1}+ \textrm{h.c.}\big) \nonumber \\
&+ \frac{1}{2}\sum_{l=\pm 1}\big( t_{\alpha}^d e^{-i (\varphi j + \varphi'_{\alpha} l)} c_{j,\alpha}^{\dagger}c_{j+l,\alpha+1}+ \textrm{h.c.}\big)\bigg],
\end{align}
where $c_{j,\alpha}$ ($c_{j,\alpha}^{\dagger}$) is the annihilation (creation) operator for a fermion on site $(j,\alpha)$. The first term is the on-site energy in the rotating frame; the second and third terms describe the nearest neighbor hopping along the real and synthetic directions, respectively; and the fourth term represents the NNN, diagonal hopping in the 2D rectangular lattice. The position dependent complex phase factor $e^{i\varphi j}$ for orbital-changing hopping results from the spatial variation of the phase of the moving lattice potential and $\varphi = 2\pi(k_{R}/k_{L})$. When a fermion hops around a unit cell, it acquires a net phase of $\Phi = \varphi+\pi$, which can be interpreted as a magnetic flux piercing through the lattice plaquette~\cite{Aidelsburger13,Miyake13}. Here the phase of $\pi$ in $\Phi$ is due to parity inversion between two intermediate orbitals. Taking into account the additional complex phase factor $e^{i \varphi_\alpha' l}$ for diagonal hopping, a subplaquette flux distribution can be assigned, as shown in Fig.~1(b), where $\phi_\alpha= \varphi/2-\varphi_{\alpha}'$. For given orbital wave functions, $t^{s}_{\alpha}/t^{d}_{\alpha}$ and $\varphi_{\alpha}'$ are determined to be functions of $k_R/k_L$ [Figs.~1(c) and 1(d)].

Our experiment starts with the preparation of a spin-polarized degenerate Fermi gas of $^{173}$Yb atoms in the $|F,m_F\rangle=|5/2,-5/2\rangle$ hyperfine spin state, as described in Ref.~\cite{Lee17}. The total atom number is $N\approx 1.0\times 10^5$ and the temperature is $T/T_F \approx 0.3$, where $T_F$ is the Fermi temperature of the trapped sample. The atoms are adiabatically loaded in a 2D optical lattice, which is formed by laser light with a wavelength of $\lambda_L=532$~nm in the $xy$ horizontal plane. The lattice spacing and depth are $a_x=\lambda_L/2$ ($a_y=\lambda_{L}/\sqrt{3}$) and $V_{L,x}=5 E_{L,x}$ ($V_{L,y}=20 E_{L,y}$) along the $x$ ($y$) direction, respectively, where $E_{L,x(y)}=\frac{\hbar^2 \pi^{2}}{2m a_{x(y)}^2}=h\times 4.1(3.1)$~kHz and $m$ is the atomic mass. Since the $y$-axis motion is frozen by the high lattice depth $V_{L,y}$ and the $z$-axis motion is irrelevant in the following experiment, our system realizes an effective 1D lattice system. Here, $k_L=\pi/a_x$ and the tunneling amplitudes are $\{t_{0}^{r},t_1^r,t_2^r\}= h \times \{0.27,1.72,3.90\}$~kHz~\cite{Heinze11}. The trapping frequencies of the overall harmonic potential are estimated to be $\{\omega_{x},\omega_{y},\omega_{z}\} \approx 2\pi \times \{64,49,135\}$~Hz. After loading the atoms in the lattice, the fractional population of the $p$ orbital is less than 6\%. 

A moving lattice potential for inter-orbital Raman coupling is generated by a pair of 556 nm laser beams propagating in the $xy$ plane, which is blue-detuned by 1.94 GHz from the $^{1}$S$_{0}$--$^{3}$P$_{1}$ transition line of $^{173}$Yb. The wave number $k_R$ of the moving lattice is determined by the $x$-axis projection of the relative wave vector of the two laser beams. The magnetic flux $\Phi$ is controlled by the laser beam arrangement, and in this work, we employ two Raman-coupling configurations, which correspond to $\Phi=1.48\pi$ and $2.44\pi$, respectively~\cite{SI}. The frequency difference $\omega$ for the two Raman beams is set to $\omega_c=[(\epsilon_1-2t_1^r)-(\epsilon_0-2t_0^r)]/\hbar$, matching the energy difference between the dispersion minima of the $s$ and $p$ bands. Under this condition, the coupling to orbitals higher than the $d$ orbital is off resonance and the system can be approximated as a three-leg ladder consisting of $s, p$, and $d$ orbitals. 

\begin{figure}[t]
\includegraphics[width=8.4cm]{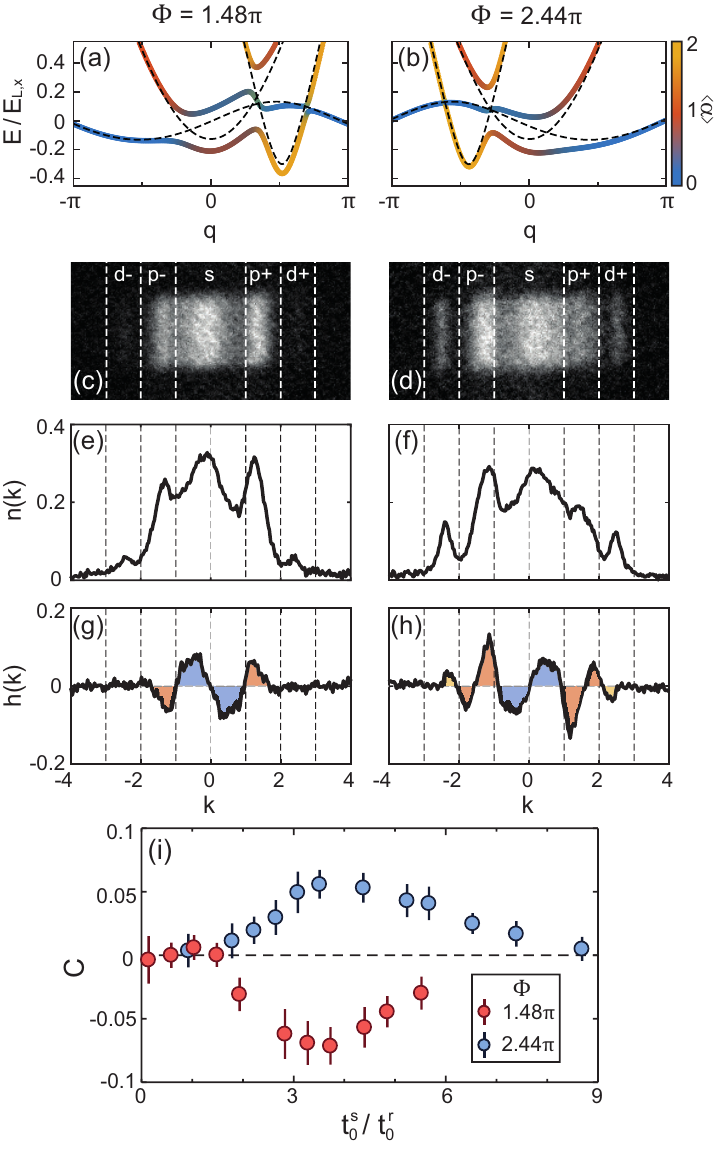}
\caption{Chiral currents in the fermionic three-leg ladder. Energy structures of the ladder system calculated from Eq.~(2) for (a) $\Phi = 1.48 \pi$  and  $t_{0}^{s}/t_{0}^{r}=3.5$, and (b) $\Phi=2.44\pi$ and $t_{0}^{s}/t_{0}^{r}=3.5$. The color indicates the mean orbital value, $\langle \alpha \rangle$. The dashed lines show the bare band structures for zero inter-leg coupling.  (c,~d) Band-mapped images of the samples adiabatically loaded into the ground band for $\Phi = 1.48 \pi$ and $2.44\pi$, respectively, and (e, f) the corresponding 1D momentum distributions $n(k)$ obtained by integrating the images along the $y$ direction. $k$ is normalized by $k_L$. (g, h) Asymmetry function $h(k)=n(k)-n(-k)$, demonstrating the chiral currents of the orbitals. (i) Evolution of $\mathcal{C}=J_0-J_1$, where $J_{\alpha}=\int^{\alpha+1}_{\alpha} h(k) dk$, as a function of $t_{0}^{s}/t_{0}^{r}$ for $\Phi=1.48 \pi$ (red) and $2.44 \pi$ (blue). Each data point was obtained by averaging twenty measurements from the same experiment, with the error bar indicating the standard deviation of the measurements.}
\end{figure}

The Bloch Hamiltonian of the three-leg ladder in momentum space is given by
\begin{equation}
H(q;\Phi) = 
\begin{pmatrix}
\epsilon_{0}(q-\Phi) &\hbar \Omega_{0}(q)/2 & 0\\
\hbar\Omega_{0}(q)/2 & \epsilon_{1}(q) -\hbar\omega & \hbar\Omega_{1}(q)/2\\
0 & \hbar\Omega_{1}(q)/2& \epsilon_{2}(q+\Phi)-2\hbar\omega
\end{pmatrix},
\end{equation}
where the $\alpha$-orbital energy dispersion is $\epsilon_{\alpha}(q)=\epsilon_{\alpha}-2 t_\alpha^r \cos (q)$ and the $\alpha$-$(\alpha+1)$ orbital coupling is $\hbar\Omega_{\alpha}(q)=t_\alpha^s-2t_\alpha^d \cos (q+\alpha \varphi-\varphi'_\alpha)$.  Here, $q$ is Bloch quasimomentum normalized by $a_x^{-1}$. Figs.~2(a) and 2(b) show the energy structures of the ladder system for our experimental conditions. The orbital-mixed ground band has a chiral region with only one pair of Fermi points, which is analogous to the chiral edge states in the integer quantum Hall effect~\cite{Kane02, Hugel14}. We note that the cross links of the ladder are manifested in the momentum dependence of $\Omega_{\alpha}(q)$. In particular, when $2t_\alpha^d > t_\alpha^s$, $\Omega_\alpha(q)$ changes its sign over a certain momentum range, implying that the topological character of the bands can change for a strong $t^d_{\alpha}$. For the two cases of  $\Phi=1.48\pi$ and $2.44\pi$,  $\{t_{0}^{s}/t_{0}^{d},t_{1}^{s}/t_{1}^{d}\}$ are estimated to be $\{4.2,1.5\}$ and $\{4.2,2.1\}$, respectively.

To probe the chirality of the ladder system, we load fermions in the orbital-mixed ground band and measure the momentum distributions of the orbitals. First, we turn on the Raman beams at the off-resonant frequency $\omega=\omega_c-2\pi\times 6$~kHz and ramp $\omega$ to the target value $\omega_c$ over 8~ms. The ramp time is limited by the scattering atom loss from the Raman beams, and in our loading process, the total atom number is reduced by 40\%. The momentum distributions of the orbitals are measured using an adiabatic band-mapping technique~\cite{Kohl05}. After suddenly turning off the Raman beams, we linearly ramp down the lattice potential to zero within 1 m and subsequently, take an absorption image of the atoms after a time-of-flight of 15 ms [Figs.~2(c) and 2(d)]. Here, the atoms in $\alpha$ band are transferred to the $\alpha$-th Brioulline zone in the momentum space of free fermions, i.e., $\alpha< |k|< \alpha+1$ ($k$ is expressed in units of $k_{L}$). Integrating the measured 2D momentum distribution along the $y$ direction, we obtain the 1D distribution $n(k)$ normalized as $\int n(k) dk =1$ [Figs.~2(e) and 2(f)].

The chiral currents of the system are clearly observed from the asymmetric momentum distributions of the orbitals. The momentum asymmetry of the $\alpha$ band is quantified with $J_{\alpha}=\int_{\alpha}^{\alpha+1}h(k)dk$, where $h(k)=n(k)-n(-k)$ [Figs.~2(g) and 2(h)]~\cite{Mancini15, Livi16}. Our measurements show   $\{J_0,J_1,J_2\}=\{-0.049, 0.020, 0.002\}$ for $\Phi=1.48\pi$ and $t_0^s= 3.3 t_0	^r$, and $\{J_0,J_1,J_2\}=\{0.037, -0.019, -0.008\}$ for $\Phi=2.44\pi$ and $t_0^s=3.5 t_0^r$.  The signs of the $J_\alpha$'s are consistent with the calculation results shown in Figs.~2(a) and 2(b). It is noticeable that the $d$-band populations are quite different in the two cases although the band structures are almost mirror-symmetric to each other. This difference originates from the different adiabaticity of the loading process due to the $q$ dependence of the inter-orbital coupling strengths. In Fig.~2(i), we display the evolution of $\mathcal{C}=J_0-J_1$ as a function of the relative inter-leg coupling, $t_0^s/t^r_{0}$.  $|\mathcal{C}|$ initially increases as $t_0^s/t^r_{0}$ increases, which is attributed to the gap opening, and reaches a maximum at $t_0^s/t^r_{0} \sim 4$ before decreasing to zero for large $t_0^s/t^r_{0}$. In the limit of $t_0^s/t^r_{0}\rightarrow \infty$, the orbital states become fully mixed to suppress the chirality of the system~\cite{Hugel14,Cornfeld15}. A similar behavior was observed in a previous experiment with a symmetric three-leg ladder system~\cite{Mancini15}.

\begin{figure}[t]
\includegraphics[width=8.4cm]{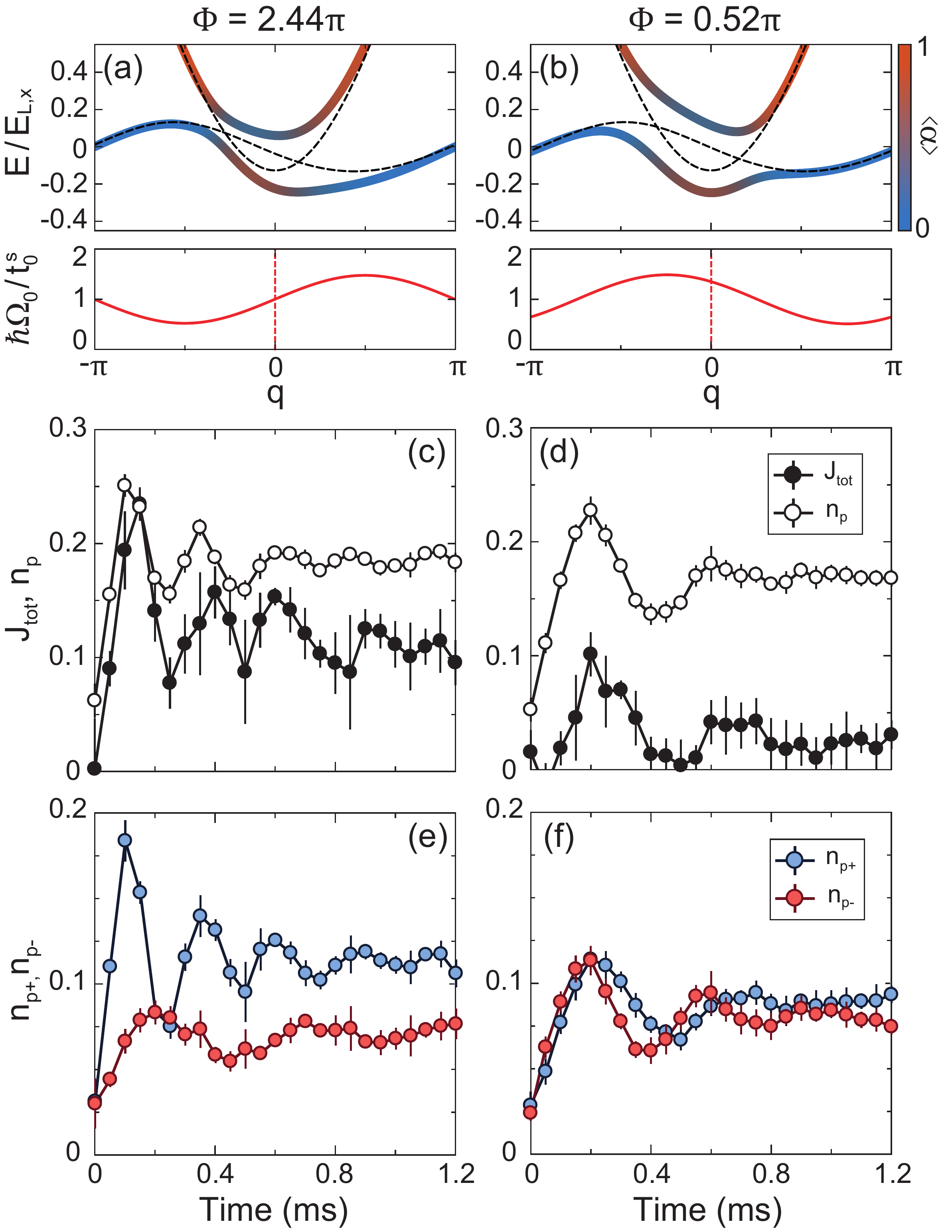}
\caption{Quench dynamics of the ladder system and effects of the complex cross links. Energy band structures of the $s$-$p$, two-leg ladder and the inter-leg coupling strength $\hbar\Omega_{0}(q)/t_0^s$ (red solid) for (a) $\Phi=2.44\pi$ and  (b) $\Phi=0.52\pi$. Quench dynamics is initiated by suddenly turning on the inter-leg coupling to the systems where atoms are initially prepared in the $s$ band. Time evolutions of the total momentum asymmetry $J_\textrm{tot}=J_{0}+J_{1}+J_{2}$ and the $p$-band fractional population  $n_{p}$ for (c) $\Phi=2.44\pi$ and  $t_{0}^{s}/t_{0}^{r}=15.2$, and (d) $\Phi=0.52\pi$ and $t_{0}^{s}/t_{0}^{r}=10.1$.  (e, f) Corresponding time evolutions of $n_{p\pm}=\int^{\pm2}_{\pm1} n(k,t)dk$. The different oscillation periods for $n_{p\pm}$ reflect the modulations of $\Omega_{0}(q)$, which originate from the complex cross links of the ladder. Each data point is the average of seven measurements, with the error bar showing the standard deviation of the measurements.}
\end{figure}

Next, we investigate the quench dynamics of the ladder system to demonstrate the effect of the cross links, wherein fermions are initially prepared in the $s$-orbital leg and the Raman beams are suddenly turned on at $\omega=\omega_c$. The sudden change of $\Omega_\alpha$ will lead to a so-called skipping cyclotron motion along the ladder edge~\cite{Mancini15}. Recalling that the inter-orbital coupling strength is modulated as $\hbar\Omega_\alpha(q)=t_\alpha^s-2t_\alpha^d \cos (q+\alpha \varphi- \varphi'_\alpha)$, we expect that the cross-link effect can be directly revealed by a momentum-resolving analysis of the quench dynamics. We examine two cases, $\Phi=2.44\pi$ and $0.52\pi$, which show almost the same effective magnetic flux $\sim\pi/2$ in a modulus of $2\pi$ but different modulation phases of $\Omega_0(q)$ with $\varphi_0'=-0.5\pi$ and $0.75\pi$, respectively [Figs.~4(a) and 4(b)]. In the case of $\Phi=2.44 \pi$ ($0.52\pi$), the average coupling strength for $q>0$, $\langle\Omega_0\rangle_+$, is stronger (weaker) than that for $q<0$, $\langle\Omega_0\rangle_-$ , so the $p$ band population with positive momentum will show faster (slower) oscillations than that with negative momentum. Here, the case of $\Phi=0.52\pi$ is generated by reversing the Raman beam directions from those for $\Phi=1.48\pi$, i.e., $k_R\rightarrow -k_R$.  

The time evolution of the total momentum asymmetry $J_{\rm tot}=\sum_{\alpha}J_{\alpha}$ and the $p$-band population $n_{p}$ are shown in Figs.~4(c) and 4(d). The in-phase oscillations for $J_{\rm tot}$ and $n_p$ are consistent with the skipping motion expected under a mangetic flux in the synthetic ladder~\cite{footnote1}. The corresponding time evolution of the $p$ band populations with positive and negative momenta, $n_{p\pm}(t)=\int_{\pm 1}^{\pm 2}n(k,t)dk$ are shown in Figs.~4(e) and 4(f). We observe that $n_{p+}$ oscillates faster (slower) than $n_{p-}$ for $\Phi =2.44 \pi$ ($\Phi=0.52\pi$), which is in agreement with that expected from the momentum dependence of $\Omega_0(q)$. The oscillation time difference is characterized by $\eta=\tau_{p-}/\tau_{p+}$, where $\tau_{p\pm}$ is the time at which the first oscillation minimum occurs in $n_{p\pm}(t)$, with our measurements giving $\eta=1.92$ and 0.81 for $\Phi=2.44\pi$ and $0.52\pi$, respectively. We find that our results are quantitatively well accounted for by the average coupling strength ratio $\langle \Omega_0\rangle_+ / \langle \Omega_0\rangle_- =(\pi t_0^s - 4 t_0^d \sin \varphi_0')/(\pi t_0^s + 4 t_0^d \sin \varphi_0')=1.9$ and 0.7 for $\Phi=2.44\pi$ and $0.52\pi$, respectively. 

\begin{figure}[t]
\includegraphics[width=8.6cm]{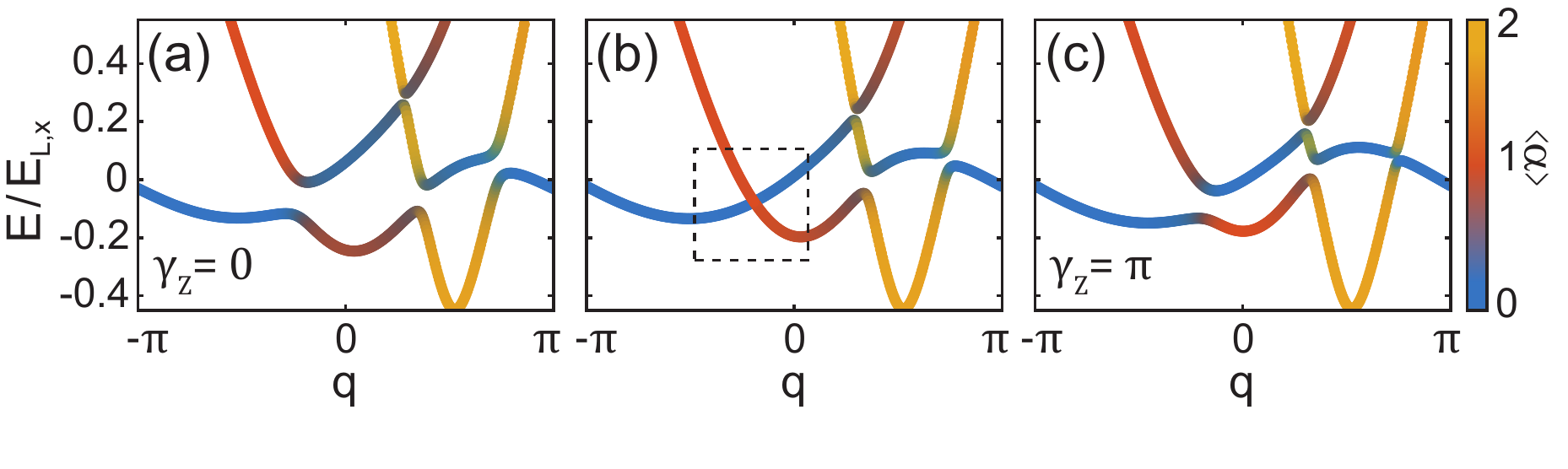}
\caption{Topological phase transition of the three-leg ladder system. Band dispersion of the system for (a) $t_{0}^{s}/t_{0}^{d}=2.5$, (b) $1.15$, and (c) 0 with $\{k_{R}/k_{L},t_0^d/t_0^r,t_{1}^{s}/t_{0}^{r},t_{1}^d/t_0^r,\varphi_{0}',\varphi_{1}'\}=\{0.24,1.5,6.1,3.6, -\pi/2,\pi/2\}$. The Zak phase of the ground band changes from $\gamma_{Z}=0$ in (a) to $\gamma_{Z}=\pi$ in (c). The topological phase transition is featured with a gap closing in the boxed region in (b).}
\end{figure}
 
It was theoretically anticipated that a chiral ladder system may undergo a topological phase transition with an increasing cross link strength~\cite{Creutz99,Hugel14,Junemann17,Sun17}. Indeed, we find that in the parameter space of our three-leg ladder system for $t_0^s/t_0^d<2$, there are multiple regions where the orbital-mixed ground band becomes topologically non-trivial with a non-zero Zak phase~(Fig.~4)~\cite{Zak89,SI}. As a means of controlling the strength and phase of the complex cross link, tailoring the orbital wave functions by engineering lattice potential is conceivable~\cite{Liu16}. We note that the $p$-$d$ orbital coupling in our ladder system shows $t_1^s/t_1^d\approx 1.5<2$ for $\Phi=1.48 \pi$, indicating the significant role of the orbital wave functions in determining the link properties.

In conclusion, we realized a cross-linked chiral fermionic ladder based on the orbital states of a 1D optical lattice. The chiral edge currents were observed and the cross-link effect was demonstrated by the momentum dependence of the inter-orbital coupling strengths. The orbital-based synthetic ladder system shows an explicitly broken leg symmetry with $t_{\alpha}^{r}\neq t_{\beta}^{r}$, providing an interesting opportunity for studying topological phases that are protected by unconventional symmetries~\cite{Song18,Hou13}. The orbital-momentum coupling scheme used in this work can be extended to multiple hyperfine spin states, which would allow for interactions between fermions~\cite{Barbarino16}.

\begin{acknowledgements} 
This work was supported by the Institute for Basic Science in Korea (Grant No. IBS-R009-D1) and the National Research Foundation of Korea (Grant Nos. NRF-2018R1A2B3003373, 2014-H1A8A1021987).
\end{acknowledgements}

\newpage


\begin{center}
\textbf{\large Supplemental Material}
\end{center}

\setcounter{equation}{0}
\setcounter{figure}{0}
\setcounter{table}{0}
\setcounter{page}{1}
\makeatletter
\renewcommand{\theequation}{S\arabic{equation}}
\renewcommand{\thefigure}{S\arabic{figure}}

\section{Multi-band tight-binding model}
In the multi-band tight-binding approximation~\cite{Dutta15}, the Hamiltonian of a particle in an one-dimensional (1D) lattice potential $V(x)=V_L \cos^2 (k_L x)$ is given by
\begin{equation}
H_0= \sum_{j,\alpha}\Big[ \epsilon_{\alpha}c_{j,\alpha}^{\dagger}c_{j,\alpha}+ \Big( (-1)^{\alpha+1}t_{\alpha}^{r}c_{j,\alpha}^{\dagger}c_{j+1,\alpha} +\textrm{h.c.} \Big)\Big],
\end{equation}
where $c_{j,\alpha}$ ($c_{j,\alpha}^\dag$) is the annihilation (creation) operator of the particle in the Wannier state $|j,\alpha\rangle$ localized at lattice site $j$ in $\alpha$ band ($\alpha=0,1,2,\cdots$ for $s,p,d,\cdots$). The on-site energy $\epsilon_{\alpha}$ and the tunneling amplitude $t^r_{\alpha}$ are given by 
\begin{eqnarray}
\epsilon_{\alpha}&=&\langle j,\alpha| \Big[ \frac{p^{2}}{2m}+V(x) \Big] |j,\alpha\rangle\\
t_{\alpha}^{r}&=& (-1)^{\alpha+1}\langle j,\alpha| \Big[ \frac{p^{2}}{2m}+V(x) \Big] |j+1,\alpha\rangle. 
\end{eqnarray}

Now we consider a situation where the particle is perturbed by a moving lattice potential, $\delta V(x,t)=V_R \cos^2 (k_{R} x -\frac{\omega t}{2})$. At the same level of approximation, the perturbations can be described by
\begin{equation}
H'= \sum_{j,l}\sum_{\alpha, \beta} \langle j,\alpha|\delta V(x,t)|j+l,\beta\rangle  c_{j,\alpha}^{\dagger}c_{j+l,\beta},
\end{equation}
where $l\in \{-1,0,1\}$. In the case that the moving lattice frequency $\omega$ is close to the band gap energy, the couplings between adjacent orbitals, i.e., $|\alpha-\beta|=1$ are most relevant and the perturbation Hamiltonian $H'$ can be further approximated as
\begin{equation}
H'= \sum_{j,\alpha}\sum_{l=0,\pm1}\Big( \langle j,\alpha|\delta V(x,t)|j+l,\alpha+1\rangle  c_{j,\alpha}^{\dagger}c_{j+l,\alpha+1} + \textrm{h.c.} \Big).
\end{equation}
Using $\langle j,\alpha| k, \beta\rangle =0$ for $\alpha\neq \beta$, the coupling amplitude can be expressed as
\begin{eqnarray}
&&\langle j,\alpha|V_{R}\cos^{2}\Big( k_{R}x-\frac{\omega t}{2}\Big) |j+l,\alpha+1\rangle \nonumber \\
&&= \frac{V_{R}}{2} \langle 0,\alpha| \cos \Big( 2k_{R}(x+ a_x j)-\omega t \Big) |l,\alpha+1\rangle \nonumber \\
&&= \frac{V_{R}}{2} \Big[ C_{\alpha}^{l}\cos(\varphi j-\omega t)-S_{\alpha}^{l} \sin(\varphi j-\omega t)\Big]
\end{eqnarray}
where $a_{x}=\pi/k_{L}$ is the lattice spacing, $\varphi = 2\pi(k_{R}/k_{L})$, $C_{\alpha}^{l}= \langle 0, \alpha|\cos(2k_{R}x)|l, \alpha+1\rangle$, and $S_{\alpha}^{l}= \langle 0,\alpha|\sin(2k_{R}x) |l, \alpha+1\rangle$ [Fig.~S1(a) and S1(b)].

In the rotating-wave approximation under unitary transformation 
\begin{equation}
U_{R}(t) = \sum_{j,\alpha} e^{i\alpha(\omega t+\frac{\pi}{2})} c_{j,\alpha}^{\dagger}c_{j,\alpha},
\end{equation}
the total Hamiltonian of the system is given by 
\begin{eqnarray}
H&=& H_0+ H' \nonumber \\
&=&\sum_{j,\alpha} \bigg[\epsilon'_{\alpha} c_{j,\alpha}^{\dagger}c_{j,\alpha}+\big( (-1)^{\alpha+1}t_{\alpha}^r c_{j,\alpha}^{\dagger}c_{j+1,\alpha} + \textrm{h.c.}\big) \nonumber \\
&&+\frac{1}{2}\big( t_{\alpha}^s e^{-i \varphi j} c_{j,\alpha}^{\dagger}c_{j,\alpha+1}+ \textrm{h.c.}\big) \nonumber \\
&&+ \frac{1}{2}\sum_{l=\pm 1}\big( t_{\alpha}^d e^{-i (\varphi j + \varphi'_{\alpha} l)} c_{j,\alpha}^{\dagger}c_{j+l,\alpha+1}+ \textrm{h.c.}\big)\bigg],
\end{eqnarray}
where $\epsilon_{\alpha}'= \epsilon_{\alpha}-\alpha\hbar\omega$, $t_{\alpha}^s =\frac{V_R}{2} S_{\alpha}^{0}$, $t_{\alpha}^d =\frac{V_R}{2}|S_{\alpha}^{1}-i C_{\alpha}^{1}|$, and $\varphi_{\alpha}'=\text{arg}(S_{\alpha}^{1}-i C_{\alpha}^{1})$. In the derivation, we used the relations of $C_{\alpha}^{0}=0$, $C_{\alpha}^{-1}=-C_{\alpha}^{1}$, and $S_{\alpha}^{-1}=S_{\alpha}^{1}$, which are easily inferred from the parity property of the Wannier states. The second and third terms represent the orbital-changing processes induced by the moving lattice potential, where the particle acquires a site-dependent phase $\varphi j$ and an additional phase $\varphi_{\alpha}'l$ depending on hopping direction and orbital in the lattice. $t_{\alpha}^{s}/t_{\alpha}^{d}$ and $\varphi_{\alpha}'$ are determined as functions of $k_R/k_L$ [Fig.~S2(c) and S2(d)].

\begin{figure}[t]
\includegraphics[width=8.4cm]{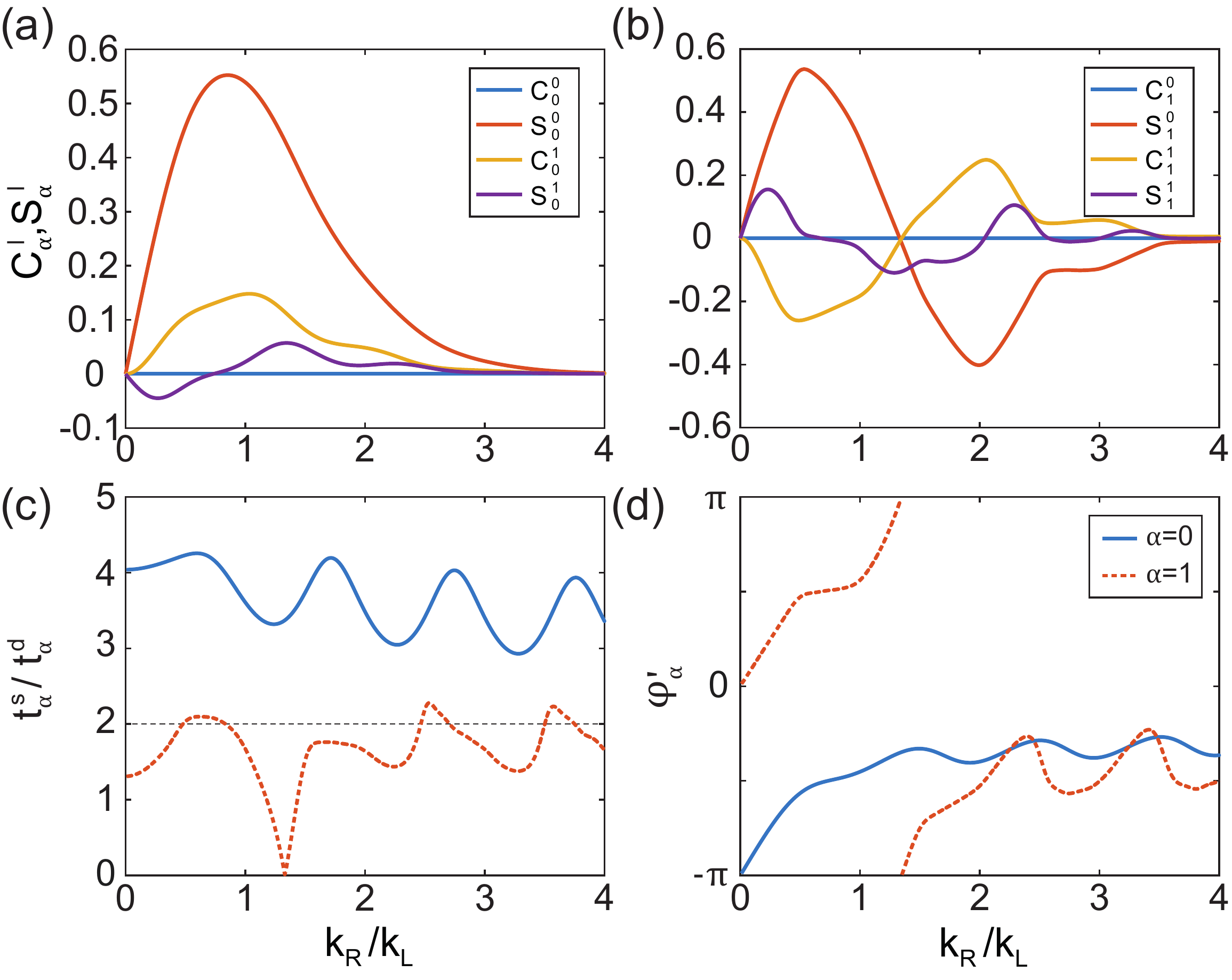}
\caption{Raman coupling coefficients $C^l_\alpha$ and $S^l_\alpha$ ($l=0,1$) as functions of $k_R/k_L$ for (a) $s$-$p$ orbital coupling ($\alpha=0$) and (b) $p$-$d$ orbital coupling ($\alpha=1$). (c, d) Corresponding $t^{s}_\alpha/t^d_\alpha$ and $\varphi_{\alpha}'$.}
\end{figure}

The momentum-space representation of the system can be obtained by the gauge and Fourier transformation 
\begin{equation}
c_{j,\alpha}=(2N)^{-1/2}\sum_{q}e^{i[(\alpha-1)\varphi+\pi]j}e^{iq j}c_{q,\alpha},
\end{equation}
where $2N$ is the number of lattice site and $q=n\frac{\pi}{N}$ ($n=-(N-1),\cdots, N$) is the quasi-momentum of the lattice. The Hamiltonian in momentum space is 
\begin{eqnarray}
H&=&\sum_{q,\alpha}\bigg[\epsilon_{\alpha}'+(-1)^{\alpha}2 t
^r_{\alpha}\text{cos}[q+(\alpha-1)\varphi]\bigg]c_{q,\alpha}^{\dagger}c_{q,\alpha} \nonumber \\
&&+\frac{1}{2}\sum_{q,\alpha}\bigg[\hbar \Omega_{\alpha}(q)c_{q,\alpha}^{\dagger}c_{q,\alpha+1}+\textrm{h.c.}\bigg],
\end{eqnarray}
where $\hbar\Omega_{\alpha}(q)=t_{\alpha}^{s}-2 t_{\alpha}^{d}\text{cos}(q+\alpha\varphi-\varphi_{\alpha}')$, which shows that the momentum dependence of the inter-orbital coupling strength results from the orbital-changing hopping process with $t_{\alpha}^{d}\neq0$.

\section{Lattice and Raman coupling setup}

The experimental setup for optical lattice and Raman coupling is sketched in Fig.~S2. A two-dimensional rectangular optical lattice is generated by using Gaussian laser beams with a wavelength of $\lambda_L=532$ nm, where a single laser beam is propagating and retro-reflected along $x$-axis, and two laser beams are placed in the $xy$ plane symmetrically with respect to the $x$-axis to intersect with each other at the angle of $2\pi/3$. The lattice constants for $x$ and $y$ directions are given by $a_x=\lambda_L/2=\pi/k_L$ and $a_y=\lambda_{L}/\sqrt{3}$, respectively. The waists of the lattice beams are $\approx 80~\mu$m and the laser beam frequency for the $y$-axis lattice is shifted by 200~MHz from that of the $x$-axis beam. In our experiment, the depth of the lattice pontential is set to be $V_{L,x}=5 E_{L,x}$ ($V_{L,y}=20 E_{L,y}$) along the $x$ ($y$) direction, where $E_{L,x(y)}=\frac{\hbar^2 \pi^{2}}{2m a_{x(y)}^2}=h\times 4.1 (3.1)$~kHz and $m$ is the atomic mass. The lattice depths were calibrated by lattice amplitude modulation spectroscopy~\cite{Heinze11}.

\begin{figure}[t]
\includegraphics[width=8.4cm]{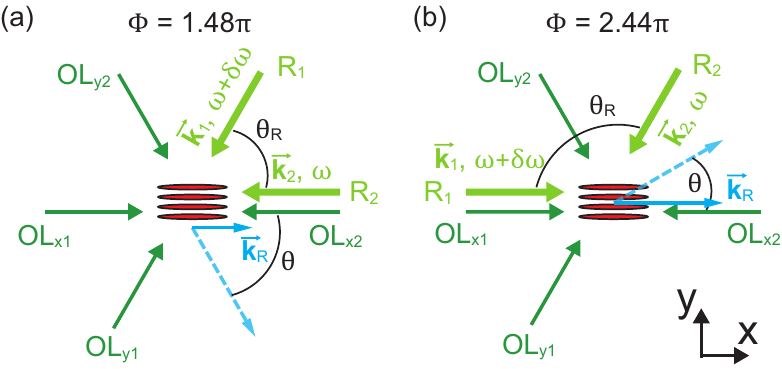}
\caption{Experimental setup for optical lattice and Raman coupling. A rectangular optical lattice is formed by two pairs of 532~nm laser beams ($\{\textrm{OL}_{x1},\textrm{OL}_{x2}\}$ and $\{\textrm{OL}_{y1},\textrm{OL}_{y2}\}$) and the inter-orbital Raman coupling is generated by two 556~nm laser beams (R$_1$ and R$_2$). Raman beam configurations for (a) $\Phi=2\pi(k_R/k_L)+\pi=1.48\pi$ and (b) $2.44\pi$.}
\end{figure}

The moving lattice potential for inter-orbital Raman coupling is generated by using two Gaussian laser beams with a wavelength of $\lambda_{R}=$556 nm, which is blue-detuned by 1.94 GHz from the $^1$S$_0$-$^3$P$_1$ narrow intercombination transition line of $^{173}$Yb. The beam waists are $150~\mu$m, much larger than the sample radius of $16~\mu$m, and the intensity variations of the laser beams over the sample are negligible. This is important to suppress the mechanical perturbations to the sample caused by the inhomogeneous AC Stark shift at a sudden turn-on of the Raman beams. In our experiment, the frequency difference $\omega$ for the two laser beams is much smaller than the band gap energy associated with the $y$-axis lattice potential and the $y$-directional Raman coupling is energetically prohibited. The wave number $k_R$ of the $x$-directional Raman coupling is given by $k_R=(\vec{k}_1 -\vec{k}_2)\cdot \hat{x} = \frac{2\pi}{\lambda_R} \sin \theta_{R} \cos \theta$, where $\vec{k}_{1,2}$ are the wave vectors of the two Raman beams, $2\theta_{R}$ is the angle between $\vec{k}_1$ and $\vec{k}_2$, and $\theta$ is the angle of $\vec{k}_1 -\vec{k}_2$ to the $x$-axis (Fig.~S2). In Fig.~S2, the two Raman-coupling configurations used in the experiment are shown, which correspond to $\Phi=k_R/k_L+\pi=1.48\pi$ and $2.44\pi$, respectively. The magnitude $t_0^s$ of the coupling between the $s$ and $p$ bands was experimentally determined by measuring the momentum-averaged Rabi frequency of band population oscillations, where the oscillations were induced by suddenly turning on the Raman beams to a sample prepared in the $s$ band. For our experimental condition, the $d$-band population was less than $8\%$ during the oscillations.

\begin{figure}[t]
\includegraphics[width=8.4cm]{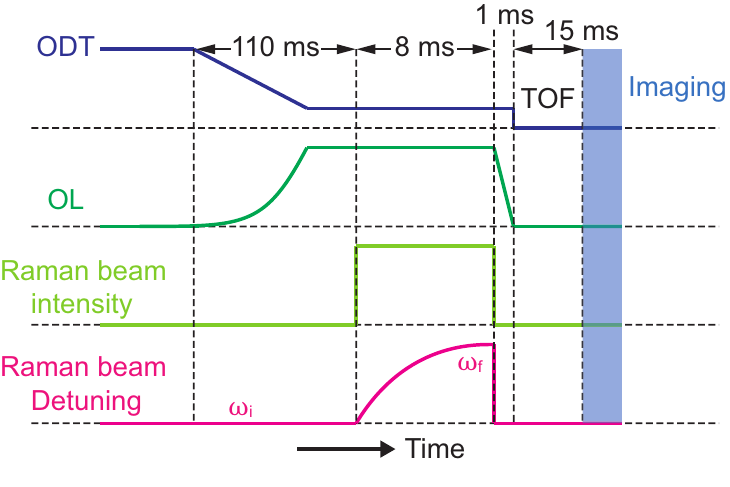}
\caption{Experimental sequence for adiabatic loading and imaging of the synthetic ladder system. The temporal control of the optical dipolt trap (ODT), the optial lattice (OL), and the Raman beam intensity and detuning are displayed. The time is not shown in absolute scale.}
\end{figure}

\section{Experimental sequence}
Figure S3 shows the experimental sequence for adiabtic loading of a Fermi gas in the orbital-momentum-coupled lattice and its imaging. First, the optical lattice is slowly ramped up to $V_{L,x}=5 E_{L,x}$ and $V_{L,y}=20E_{L,y}$ over 110~ms. During the lattice turn-on, a magnetic field is applied to the sample and increased up to $\approx 153$~G, inducing a Zeeman energy splitting of $\approx h\times 34$~kHz between neighboring hyperfine ground states. This is to prevent unwanted hyperfine state-changing Raman transitions. Then, the Raman beams are switched on and their frequency difference $\omega$ is exponentially changed from $\omega_c-2\pi\times 6$ kHz to the target frequency $\omega_c$ over 8~ms. Because of the scattering atom loss from the Raman beams, the total atom number was reduced by $\approx 40\%$ in the loading process. 

To measure the lattice momentum distributions of the synthetic ladder system, a band-mapping technique is employed~\cite{Kohl05}, where the Raman beams are suddenly turned off, then the lattice potential is ramped down to zero within 1~ms, and finally the optical trapping potential is switched off. In this process, the lattice system is slowly mapped to a free particle system by transforming the orbital states into the free-space momentum states in the corresponding Brioulline zones. The momentum distribution of the atoms is measured by taking an absorption image after a time-of-flight of 15~ms using the $^{1}$S$_{0}$-$^{1}$P$_{1}$ atomic transition.  

\begin{figure}[t]
	\includegraphics[width=7.0cm]{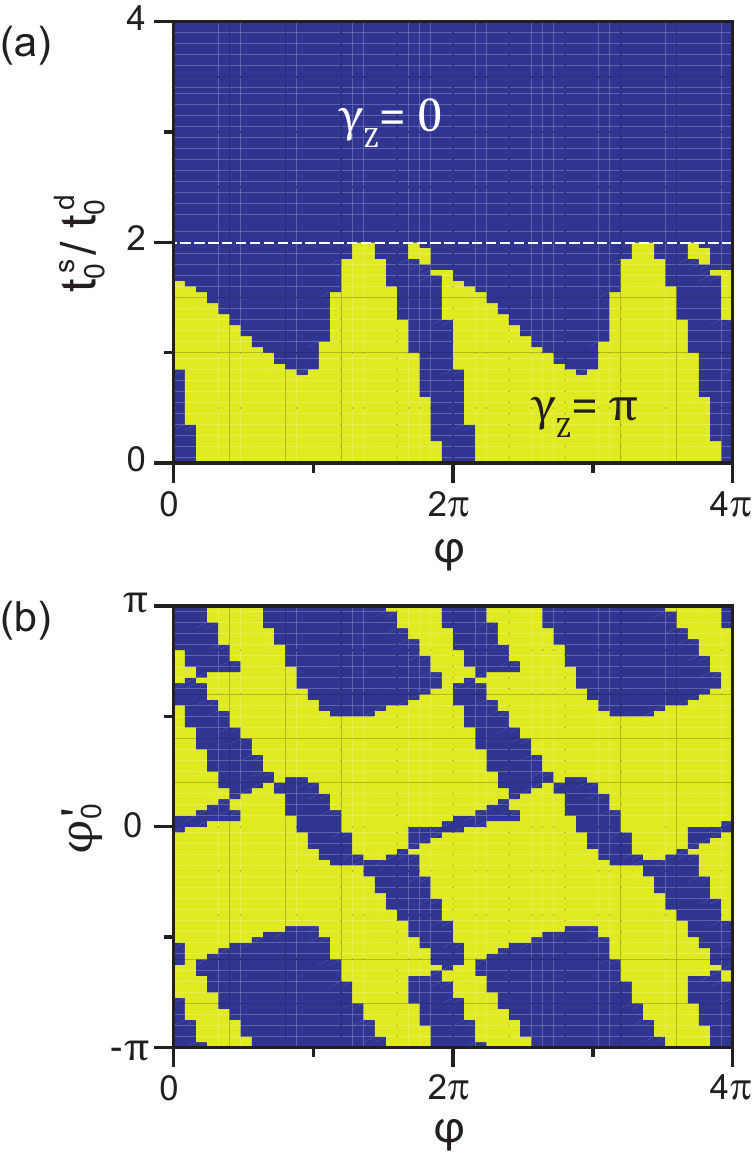}
	\caption{Zak phase $\gamma_Z$ of the ground band of the three-leg ladder system (a) in the plane of $\varphi$ and $t^s_0 / t^d_0$ for $\{t_{1}^{s}/t_{1}^{d},\varphi_{0}',\varphi_{1}'\}=\{1.7,-\pi/2,\pi/2\}$, and (b) in the plane of $\varphi$ and $\varphi'_0$  for $\{t_{0}^{s}/t_{0}^{d},t_{1}^{s}/t_{1}^{d},\varphi_{1}'\}=\{1,1.7,\pi/2\}$. The yellow regions indicate the topological regimes with $\gamma_{Z}=\pi$. The white dashed line in (a) indicates the line with $t_{0}^{s}/t_{0}^{d}=2$. In the experimental condition for $\Phi=1.44\pi$ ($\varphi=0.44\pi$), $\{t_{0}^{s}/t_{0}^{d}, t_{1}^{s}/t_{1}^{d},\varphi_{0}',\varphi_{1}'\}=\{4.2,1.5,-0.75\pi,0.24\pi\}$. }
\end{figure}

\section{Topological phases}

The three-leg ladder model described by $H(q;\Phi)$ in Eq.~(2) constitutes an 1D three-band system and its topological property can be characterized by the Zak phases of the bands~\cite{Zak89}. The Zak phase is defined by  
\begin{equation}
\gamma_{Z} = i\int_{BZ} \langle u_{q}^{n}|\partial_{q}|u_{q}^{n}\rangle dq,
\end{equation}
where $u_{q}^{n}$ is the cell-periodic Bloch function of the $n$-th band and by numerically calculating $\gamma_Z$ of, e.g., the orbital-mixed ground band, we may investigate on the condition for the three-leg ladder system to have a topologically non-trivial phase. Some of the calculation results for a parameter region close to our experimental condition are displayed in Fig.~S4. We find that a topological  phase with $\gamma_Z\neq0$ can emerge in the system for the variations of the complex cross-link amplitudes. In particular, we observe that the topologically non-trivial phase exists only in the regions with $t_{\alpha}^{s}/t_{\alpha}^{d}<2$. This is consistent with our anticipation from the relation of $\hbar\Omega_{\alpha}(q)=t_\alpha^s-2t_\alpha^d \cos (q+\alpha \varphi-\varphi'_\alpha)$, where the inter-band coupling amplitude changes its sign in a certain range of $q$ for $t_{\alpha}^s/t_{\alpha}^d <2$. 

In a two-band model such as the Creutz ladder model~\cite{Creutz99}, it can be easily understood that a topologically non-trivial band with $\gamma_Z\neq 0$ will appear when the inter-band coupling amplitude changes its sign at the two distinct band crossing points because it means that the spin texture of the mixed band has a full winding in the pseudo-spin space composed by the two band states. In a similar manner, we expect that the topological property of the three-leg ladder system would be determined by the sign changes of two coupling amplitudes $\Omega_{0}(q)$ and $\Omega_{1}(q)$, and their relative positions to the band crossing points which are determined by $\varphi$ and $\varphi'_\alpha$. In Fig.~S4(b), for example, we see that the Zak phase depends on $\varphi$ and $\varphi'_0$ even for fixed $t^{s}_\alpha/t^{d}_\alpha <2$. It would be interesting to clarify the geometric meaning of having nonzero $\gamma_z$ in terms of the sign change of $\Omega_\alpha$ and the band crossing points in the pseudo-spin-1 system.


\begin{references}
\bibitem{footnote1} The dephasing time scale of oscillation was about 1~ms, which is attributed to the averaged dynamics from different initial states~\cite{Mancini15} and fluctuations of the relative phase of the Raman beam pair.
\end{references}
\end{document}